\documentclass[]{aa}
\usepackage[varg]{txfonts}
\usepackage[utf8]{inputenc}
\usepackage{amsmath}
\usepackage{amsfonts}
\usepackage{amssymb}
\usepackage{graphicx}

\begin{document}
\title{Principal Components of Short-term Variability in Venus' UV Albedo}
\titlerunning{Venus Albedo Components}
\authorrunning{Kopparla et al.}
	\author{Pushkar Kopparla \inst{1,2}
    \and Yeon Joo Lee\inst{1} 
    \and Takeshi Imamura\inst{1}
    \and Atsushi Yamazaki\inst{3}} 

	\offprints{P. Kopparla, \email{pushkarkopparla@gmail.com}}
	
	\institute{Department of Complexity Science and Engineering, Graduate School of Frontier Sciences, The University of Tokyo, Kashiwa, Chiba Prefecture, Japan 277-8561
  \and JSPS International Research Fellow 
  \and Institute of Space and Astronautical Science (ISAS), Japan Aerospace Exploration Agency (JAXA), Sagamihara, Kanagawa Prefecture, Japan 252-5210}
\abstract{We explore the dominant modes of variability in the observed albedo at the cloud tops of Venus using the Akatsuki UVI 283-nm and 365-nm observations, which are sensitive to SO$_2$ and unknown UV absorber distributions respectively, over the period Dec 2016 to May 2018. The observations consist of images of the dayside of Venus, most often observed at intervals of 2 hours, but interspersed with longer gaps. The orbit of the spacecraft does not allow for continuous observation of the full dayside, and the unobserved regions cause significant gaps in the datasets. Each dataset is subdivided into three subsets for three observing periods, the unobserved data are interpolated and each subset is then subjected to a principal component analysis (PCA) to find six oscillating patterns in the albedo. Principal components in all three periods show similar morphologies at 283-nm but are much more variable at 365-nm. Some spatial patterns and the time scales of these modes correspond to well-known physical processes in the atmosphere of Venus such as the $\sim$4 day Kelvin wave, 5 day Rossby waves and the overturning circulation, while others defy a simple explanation. We also a find a hemispheric mode that is not well understood and discuss its implications.}

\maketitle

\section{Introduction}
The atmosphere of Venus, as seen in ultraviolet wavelengths, is long known to have striking albedo patterns that are indicative of the dynamics and chemistry of the upper atmosphere \citep{ross1928photographs,del1990planetary,markiewicz2007morphology}. Several of these features have been identified and named, for instance: the Y-feature, associated with equatorial Kelvin wave \citep{del1990planetary,peralta2015venus}; mid-latitude Rossby waves \citep{del1990planetary}; the equatorial and polar caps and bands \citep{rossow1980cloud}. The main absorbers in the near UV, SO$_2$ and the unknown UV absorber, seem to vary on multiple timescales, from daily to multi-year \citep{encrenaz2012hdo,marcq2013variations,lee2015long} and their generation, destruction and advection in the atmospheric flow are responsible for the albedo patterns. 

The $\sim 4-5$ Earth day period variations are known to be related to the atmospheric background flow and short period Kelvin and Rossby waves \citep{khatuntsev2013cloud,kouyama2015vertical,peralta2015venus}, but not fully understood since the relationship between the UV albedo contrast and the physical quantities describing the wave field (e.g., temperature, pressure, velocity) is not known. The causes for the longer period variability are uncertain, such as 255 days variability in zonal wind speed \citep{kouyama2013long}, and $\sim$270 days in the 365-nm latitudinal contrast \citep{lee2015long} and mesospheric SO$_2$ gaseous abundance \citep{marcq2013variations}. With sustained observations over a 3 year period now available from the Akatsuki Ultraviolet Imager (UVI) instrument \citep{nakamura2016akatsuki,yamazaki2018ultraviolet}, we have a long enough baseline of observations to find the leading modes of oscillation in the albedo, their periodicities and associated spatial structures. Using observations at 283-nm, which are correlated to SO$_2$ abundance above the clouds (with some absorption by unknown UV absorber and ozone) and at 365-nm (close to the peak absorption by the unknown UV absorber), we attempt to answer two main questions in this paper:
\begin{enumerate}
\item Can we resolve the variable UV patterns into a small number of recognizable components?
\item What do these components tell us about the physical processes active in Venus' atmosphere?
\end{enumerate}

The next section details the data reduction, including the handling of missing data  and principal component analysis. The following section deals with the leading oscillations that result from the PCA, their physical interpretations and the statistical significance of the results against noise. The final section summarizes the findings and speculates on future applications of such analyses for Venus climate studies.
\section{Data and Methods}
\subsection{Data Reduction and Normalization} \label{ssec:datareduction}

\begin{figure*}
  \includegraphics[width=\linewidth]{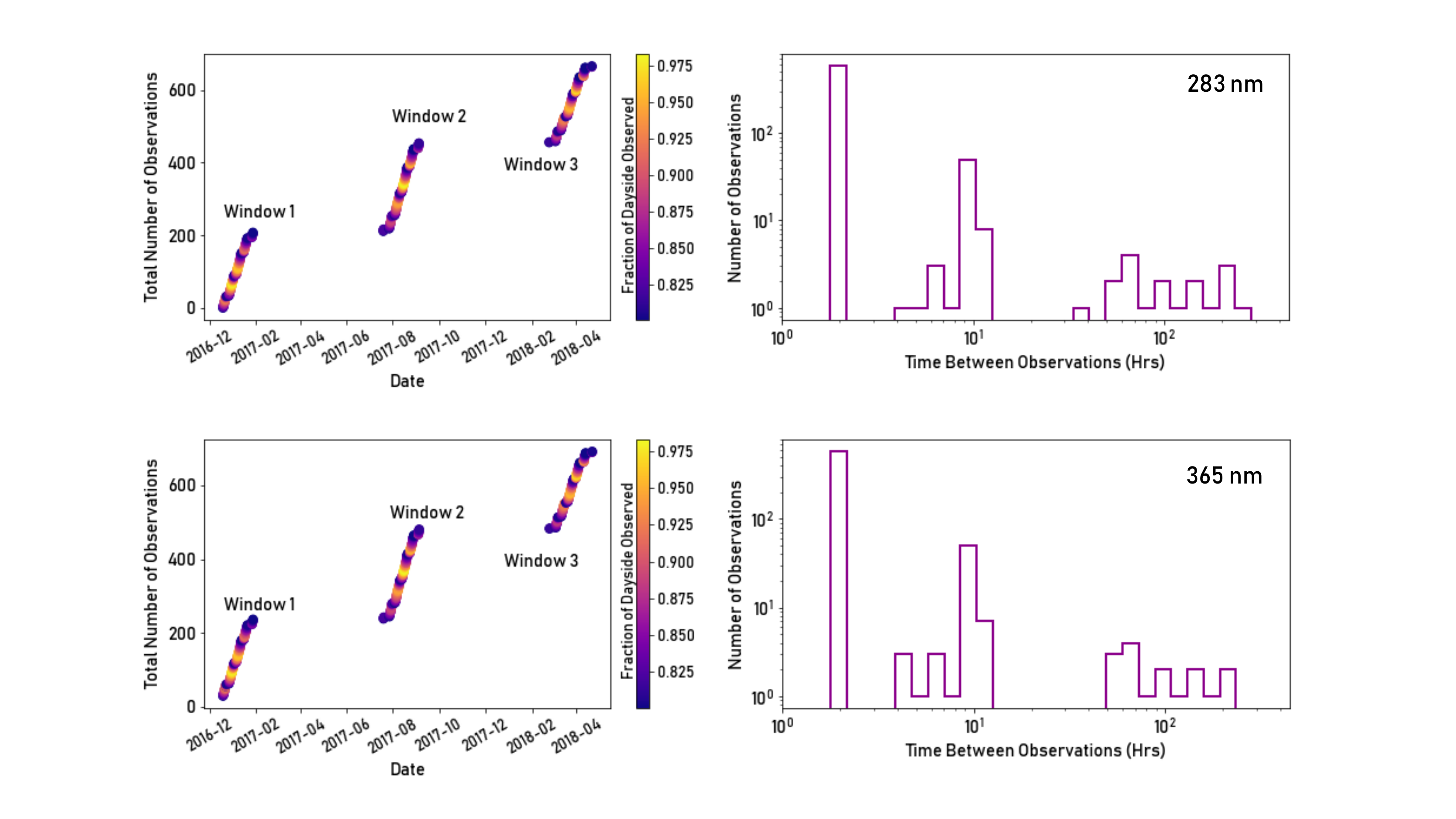}
  \caption{Plots showing observation times of the 665 and 652 images used in this study for 283-nm [top left] and 365-nm [bottom left] respectively. The windows consist of 209, 244 and 212 images for 283-nm and 206, 237 and 209 images for 365-nm respectively. The color of the data point indicates the fraction of the dayside that was observed in that image. [Right] Histograms showing the time gaps between successive images. The majority of the images are taken at intervals of 2 hours, but longer intervals are not uncommon.}
  \label{fig:datadistribution}
\end{figure*}

The Ultra Violet Imager (UVI) on Akatsuki has two filters at 283 and 365-nm, which correspond to the absorption bands of SO$_2$ and the unknown UV absorber. The field-of-view (FOV) is $12^\circ \times 12^\circ$ can observe the whole Venus disk except for about 8 hours near the periapsis. The imaging area is composed of 1024x1024 pixels, with an angular resolution of $12\times10^{-2}$ degrees per pixel, which corresponds to spatial resolutions of about 200 m and 86 km on the cloud top level in the observations from the altitudes of $\sim$1000 km at the periapsis and 60 Venus radii at the apoapsis \citep{yamazaki2018ultraviolet}. We use the Level-3 (L3) 283-nm data from the UVI, internal release version  20180901. The L3 data from Akatsuki have observed variables mapped onto a regular (equi-spaced) longitude–latitude grid.  The resolution of the L3 longitude–latitude map is $0.125^\circ \times 0.125^\circ$ $(2880 \times 1440)$ grids for $360^\circ$ longitude and $180^\circ$ latitude) \citep{ogohara2017overview}. The orbit of the spacecraft is elliptical with a revolution period of $\sim$10.5 Earth days, a periapsis altitude of 1000–8000 km and an apoapsis altitude of $\sim$360,000 km \citep{nakamura2016akatsuki}. The data used in this study were taken on the apoapsis side, where the FOV of UVI  can capture the whole Venus disk. Since the major axis of the orbit is roughly fixed in the inertial coordinate, the observation condition changes with the revolution of Venus around the Sun. A visualization of the dataset is provided in Fig \ref{fig:datadistribution}. 

To identify spatial structures and time series of oscillations, it is ideal to have continuous observations of the full dayside at high frequency over a long time period. However, due to practical constraints, the fraction of dayside observed varies systematically and there are often long gaps in the time series. The majority of observations have a gap of 2 hours between them, though the longest gap between successive observations is about 40 days which is the time of missing global day side images due to the highly elliptical equatorial orbit. Images that cover less than $80\%$ of the dayside observed are excluded from this analysis and grid points on a given image where the dayside is not observed are referred to as missing data points. The dayside is the region between solar local time 600-1800 hrs, and the fraction of the dayside observed is calculated as the ratio of grid points in this region with observed values to the total number of grid points on the dayside. The exclusion threshold could be relaxed to $10-60\%$, but doing so increased the ratio of missing data to observed data and introduced too many artificial patterns during the interpolation process described in Sec \ref{sssec:dineof}. On the other hand, as the exclusion threshold is increased, the number of images that can be included in the analysis decrease. As a result, the statistics become more uncertain and the time baseline shorter, limiting the longest period oscillation which can be studied. The $80\%$ threshold offers a reasonable balance between these two competing effects.

We convert the radiances to relative albedos based on the Minnaert law described in \citet{lee2017scattering}, using eq. 8 and 10 of that paper for 283-nm and 365-nm respectively. The albedos in each observation thus derived are normalized by dividing by the spatial mean value in each map to give a relative albedo. We have chosen to normalize by the spatial mean instead of the commonly used spatial maximum, since the maximum is very unstable in fields with a large number of missing values. Alternately, another stable measure of the top values of the albedo distribution (such as the 90$^{th}$ percentile) could be used for normalization. Note that this normalization allows for values of albedo above 1 and removes trends in albedo with time, but that is not our primary concern at this time. We are interested in planetary scale patterns in albedo, for which the relative albedos are sufficient. The maps are then centered such that the sub-solar longitude is always at $180^\circ$ or 1200 local time. This is equivalent to plotting on a local solar time grid. Grid points where the cosines of the solar zenith angle or the viewer zenith angle are $\leq$0.2 are excluded, since the observing error at such points is rather large compared to natural variability \citep{fukuhara2017large}. The data is regridded to a $2^\circ$ resolution (using the \texttt{resize} function from the python library \texttt{skimage.transform}). The region of interest for our study is the dayside, which we define to be between 800-1600 hrs local time. The 283-nm dataset contains 665 images, each with 4476 grid points on the dayside. Of these 2985492 points, $8\%$ are unobserved ("missing data"). The 365-nm dataset contains 652 images, with a similar fraction of missing data.
\subsection{Principal Component Analysis}
PCA is a powerful dimensionality reduction technique, and is widely used to find oscillations in atmospheric and climate data on Earth \citep{wilks2006statistical}. Our PCA will follow this procedure:
\begin{enumerate}
\item The two dimensional (latitude longitude grid) images containing relative albedos are flattened into one dimensional arrays.
\item A data matrix is created with $p$ variables (here the number of spatial gridpoints, 4476) as rows and $n$ observations (here the number of images) as columns. An element $x_{i,j}$ of this matrix corresponds to the $i^{th}$ gridpoint and $j^{th}$ observation.
\item The time mean (row-wise mean) of each variable is subtracted to generate the anomalies from the mean.
\begin{equation}
x^{a}_{i,j} = x_{i,j}-\frac{\sum^n_{j=1}x_{i,j}}{n} = x_{i,j}-\bar{x_i}
\end{equation}
\item The covariance matrix is calculated with the dimension $p*p$. An element of this matrix $c_{k,l}$ is the covariance between the observations at gridpoints $k$ and $l$, which is calculated as
\begin{equation}
c_{k,l} = \frac{\sum^n_{j=1} x^{a}_{k,j} x^{a}_{l,j}}{n-1}
\end{equation}
\item The eigenvalues and eigenvectors of the covariance matrix are calculated (we use the \texttt{eigsh} function from the python library \texttt{scipy.sparse.linalg}). The eigenvectors are called the principal components (PCs) or empirical orthogonal functions (EOFs) which give the spatial structure of the leading modes of variability. Each PC has a dimension of $p*1$ and there are $p$ PCs in total.
\item The dot product of the mean-removed data with the PCs give the principal component loadings (hereafter referred to as loadings). The loadings give the time series of contribution of each mode of variability to the data, each of which has a shape $1*n$ and there are $p$ loadings in total. The loadings are derived from the EOFs, $\epsilon$, and data anomalies, $x^a$ in this way:
\begin{equation}
L_{k,j} = \sum^n_{i=1}\epsilon_{k,i}\,.\,x^a_{i,j}
\end{equation}
\item The original data can be fully reconstructed from the PCs and loadings thus:
\begin{equation}
x_{i,j} = \bar{x_i} + \sum_{k=1}^p \epsilon_{k,i} L_{k,j}
\end{equation}
\end{enumerate}
The PCs corresponding to the largest few eigenvalues and their loadings give the spatial patterns and the time series of the leading oscillations of the dataset, respectively. The fraction of total data variance explained by each eigenvector is given by the ratio of its eigenvalue to the sum of all eigenvalues. 
\subsubsection{Calculation of the Covariance Matrix}

The covariance of albedo values between two pixels (true covariance or population covariance) has to be approximated using the limited data sample available (sample covariance). However, the existence of missing data means that the sample covariance matrix cannot be calculated immediately from the data. The simplest approaches for dealing with missing data in the calculation of the sample covariance matrix are deletions - removing variables with missing data points (known as list-wise deletion) or the calculation of covariances by considering only data points where data exist for both variables of interest (pair-wise deletion) \citep[for e.g.,][]{carter2006solutions,nakagawa2008missing}. List-wise deletion is undesirable, since it discards a grid point entirely even if that point has missing data at a single time step only. Also, use of pair-wise deletion introduces undesirable properties to the sample covariance matrix, such as the existence of negative eigenvalues. This indicates that the sample covariance is not a good estimate of the true covariance of the system being observed and the sample covariance matrix lacks the properties of a true covariance matrix such as positive semi-definiteness (only positive or zero eigenvalues) \citep{pourahmadi2011covariance}. Using such bad estimates often results in the first few eigenvalues being biased high. Meanwhile, existence of negative eigenvalues makes the calculation of the variance associated with each eigenvalue hard to interpret, since negative variances do not have a clear physical meaning \citep{dray2015principal}. 

A better way to deal with missing values is imputation, either by interpolation from available data, or the imposition of additional constraints on the structure of the covariance matrix (regularization). Several different techniques have been proposed to regularise the properties of such ill-conditioned sample covariance matrices such as ridge regressions/Tikhonov regularization, banding, tapering \citep{bickel2008regularized,warton2008penalized}. These methods are primarily focused on eliminating bad behavior in the covariance matrix by introducing constraints such as smoothness or sparsity, where the constraints may or may not be determined from the data set. Other approaches, often used in climate studies, fill in missing data through various interpolations - nearest neighbor regressions followed by smoothing [DCT-PLS] \citep{garcia2010robust,wang2012three}, optimal interpolation \citep{burgess1980optimal}, singular spectral analysis \citep{kondrashov2006spatio} or iterative techniques like regularized estimation maximization [RegEM] \citep{schneider2001analysis}, data interpolating empirical orthogonal functions [DINEOF] \citep{beckers2003eof} and several other such variants of PCA based interpolations \citep{ilin2010practical}. For our dataset, we need an interpolation technique with the following properties:
\begin{enumerate}
\item The interpolation method must be able to handle large fractions of missing data and continuous data gaps, and must be able to effectively interpolate in three dimensions (as opposed to purely spatial interpolation). It must derive the regression parameters from the data itself, i.e., be non-parametric.
\item Since the missing values lie on a smooth map, there is a roughness constraint on the interpolated values. This can be enforced in many ways: creating interpolations using low-order principal component truncations (like DINEOF), ridge regularizing the covariance matrix (RegEM) or enforcing a roughness penalty by explicitly specifying a smoothness parameter (DCT-PLS).
\end{enumerate}
Since our ultimate goal is to perform a PCA on this dataset, two of the above methods, RegEM and DINEOF, iteratively create imputed datasets that converge on a set of PCs. But RegEM uses a 2d spatial-only linear regression which is unsuited for our data (since missing values are primarily located near the edges of the region of interest rather than in the center), so we use the DINEOF method \citep{beckers2003eof}. It is closely related to other spatio-temporal methods like singular spectral analysis and optimal interpolation and produces comparable results \citep{allen1996monte, alvera2005reconstruction}.
\subsubsection{Data Interpolating Empirical Orthogonal Functions}
\label{sssec:dineof}
 The DINEOF method is based on the descriptions from \cite{beckers2003eof,alvera2005reconstruction} and is implemented as follows:
\begin{enumerate}
\item All missing values in the time-mean subtracted dataset are filled in with zeros and the covariance matrix is calculated.
\item The PCs, eigenvalues and loadings of this zero filled covariance matrix are calculated as described in Sec \ref{ssec:datareduction}. 

\item The missing values are imputed using a reconstruction consisting only of the first leading PC and corresponding loadings like this:
\begin{equation}
x^a_{i,j} =  \epsilon_{1,i} L_{1,j}
\end{equation}

\item The imputed dataset is again used to calculate a new covariance matrix, and the previous step is repeated to find better imputation values. This procedure is iterated till the imputed values converge. Convergence is defined to occur when the root mean square difference in imputed albedo values from two successive iterations differ by less than $10^{-4}$ or 10$\%$ over three successive iterations, computed thus.

\begin{equation}
 RMS_{(n,n-1)} \leq 10^{-4}~~ or ~~ \frac{|RMS_{(n,n-1)}-RMS_{(n-1,n-2)}|}{RMS_{(n-1,n-2)}} \leq 0.1
\end{equation}
where RMS$_{(n,n-1)}$ is the root mean square difference in imputed values between iterations $n$ and $n-1$.
\item The next set of iterations then uses two PCs for the imputation:
\begin{equation}
x^a_{i,j} =  \epsilon_{1,i} L_{1,j} + \epsilon_{2,i} L_{2,j}
\end{equation}
and is repeated to convergence. 

\item Thus we can arrive at a converged imputed dataset for any given number of eigenvectors.
\end{enumerate}

Since it is possible to create an imputed dataset with any $k$ PCs, where $1 \leq k \leq p$, the optimal number of PCs is determined through a generalized cross validation procedure.  Following \cite{alvera2005reconstruction}, we choose $1\%$ of the existing data points randomly and set them as artificially missing. This number of points is often used for cross-validation procedures \citep[for e.g.,][]{wilks2006statistical}. The imputed datasets are reconstructed using 1,2,3 .. 100 PCs using the convergence criterion above. The artificially missing data is not included in the convergence calculation as described in Step 4 above. For each converged imputed dataset, the root mean square error of the reconstruction of the artificially missing data is calculated. The optimal imputation is the one for which this RMS error is minimum. For our data set, this minimum occurs at $N\sim 25$ PCs. 
\begin{figure*}
  \includegraphics[width=\linewidth]{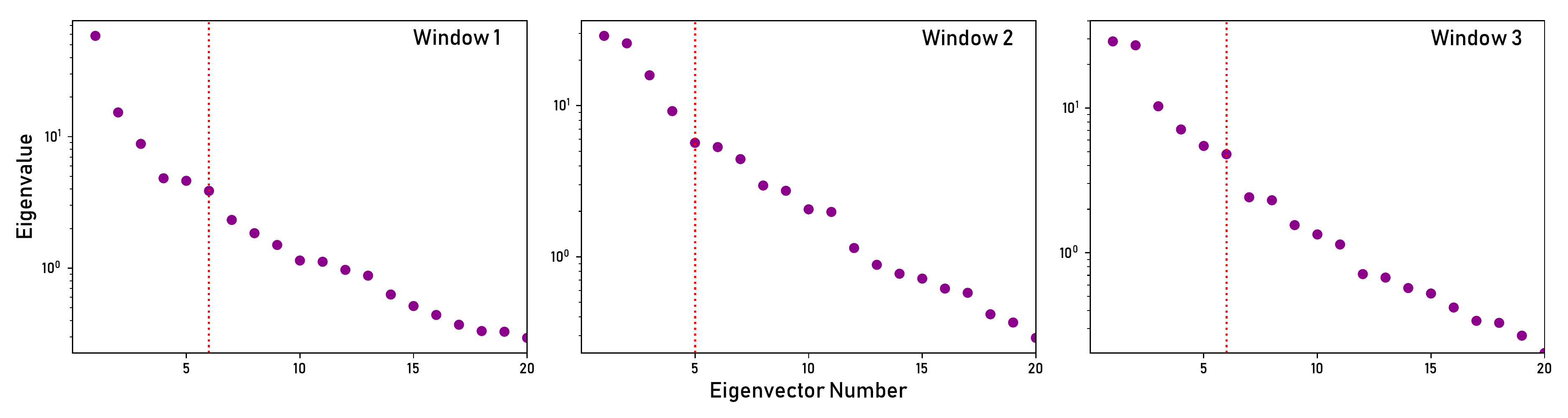}
  \caption{Plot showing the distribution of the twenty largest eigenvalues in the DINEOF interpolated datasets for 283-nm. The 365-nm datasets show similar trends. The red vertical line shows the point at which the eigenvalue curve has an inflection. PCs corresponding to eigenvalues left of this line are considered significant by the Scree test.}
  \label{fig:covproperties}
\end{figure*}

\section{Results and Discussion}
\subsection{Dominant Modes and Statistical Significance}
The question of how many PCs are significant and should be retained has been widely discussed, and many different stastical and rule-of-thumb approaches exist \citep{jackson1993stopping,peres2005many,cangelosi2007component}. Of these we use a very simple criterion that looks for an inflection point in the eigenvalue spectrum to find the break between signficant and noisy PCs. This is known as Catell's Scree test \citep{cattell1966scree}, and is a commonly used heuristic significance metric. From Fig \ref{fig:covproperties}, we see that our interpolated datasets have an inflection point around 5 or 6, indicating a dimension of 6. We therefore retain only the first 6 spatial PCs, and these are shown in Figs \ref{fig:principalcomps} and \ref{fig:principalcomps2}. 

\begin{figure*}
\centering
  \includegraphics[width=\linewidth]{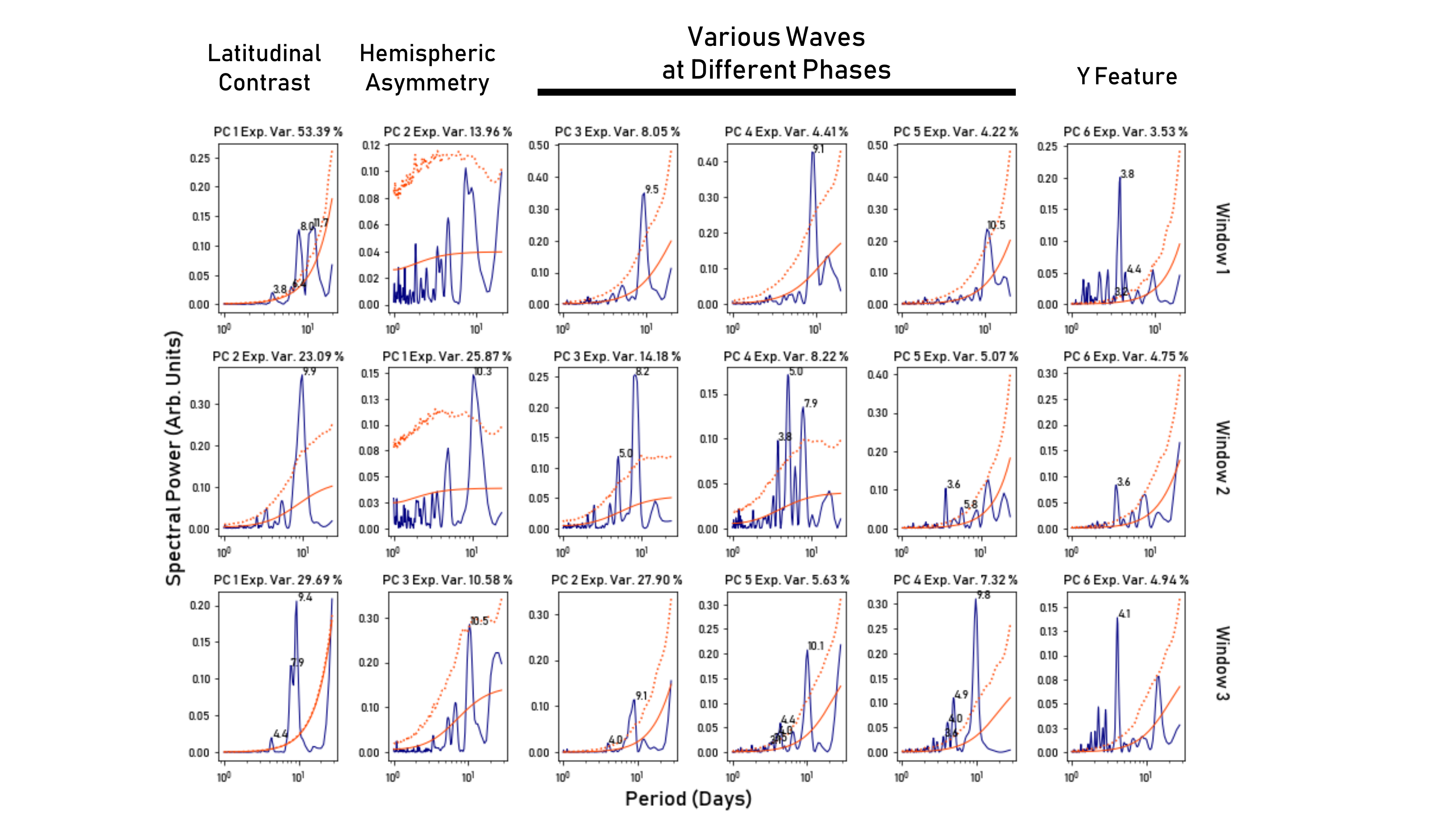}
  \caption{Lomb-Scargle periodograms for the loadings corresponding to the first six PCs for three windows at 283-nm. The numbers given for the spectral peaks indicate the associated periods. The solid red line is the average spectral power of red noise, while the dotted red line shows the $95\%$ confidence limit of the distribution from 1000 Monte Carlo AR1 red noise processes. The explained variance fraction listed in the title of each subplot is calculated as the ratio of the mode's eigenvalue to the sum of all eigenvalues from the covariance matrix, expressed as a percentage.}
  \label{fig:timeseries}
\end{figure*}

The Lomb-Scargle periodograms of the time series corresponding to each PC are shown in Figs  \ref{fig:timeseries} and \ref{fig:timeseries2}, with a minimum period of 1 day and a maximum of half the total length of the window (typically each window is about 40 days). The periodograms show several peaks, but their significance must be first confirmed. The spectral analyses of atmospheric data are affected by red noise \citep[for e.g.,][]{allen1996monte,meinke2005rainfall}, where the noise contribution increases with period as opposed to a flat spectrum for white noise. This leads to spurious long period peaks in periodograms. This noise is usually approximated by an autoregressive process with a time lag of one (AR1) \citep{gilman1963power}, however irregular time sampling prevents estimation of the autocorrelation coefficient directly from the data. We therefore use existing methods in the climate science literature \citep{mudelsee2002tauest,schulz2002redfit} for variable spaced time series data. We do not use the programs made available by these studies, but only use the methods as described in the text with some minor modifications. Our implementation is summarized below:
\begin{enumerate}
\item The AR1 process is modeled as 
\begin{equation}
x\left(t\right) = x\left(t-1\right)exp\left(-\frac{\Delta t}{\tau}\right) + \epsilon
\end{equation}
\newline
where $x\left(t\right)$ is the time series value at time $t$, $\Delta t = \left(t\right)-\left(t-1\right)$ is the time step, $\tau$ is persistence timescale of the time series and $\epsilon$ is the Gaussian noise component.
\item $\tau$ is estimated by a least-squares fit of the AR1 equation to the loadings (using the function \texttt{curve\_fit} from the python library \texttt{scipy}). There is one $\tau$ for each principal component loading in each window.
\item 1000 artificial time series are generated using the AR1 process described above with the initial value and the time steps being the same as the PC loading. The Gaussian process is taken to have a variance of 
\begin{equation}
\sigma^2 = 1 - exp\left(-\frac{2\Delta t}{\tau} \right)
\end{equation}
\item Each artificial time series is scaled so that the periodogram has the same area under the curve as the original time series. The mean, $<G_{am}(f)>$, and the standard deviation, $<G_{as}(f)>$ , of the set of all the artifical time series' periodograms are calculated, where $f$ is the frequency.
\item The analytical expression for the power spectrum in the periodogram of an AR1 process is given by 
\begin{equation}
G(f) = G_o \frac{1-\rho^2}{1-2\rho*cos(\pi f/f_{max})+\rho^2}
\end{equation}
where $f_{max}$ is the highest frequency in the periodogram (often taken to be the Nyquist frequency, but here set to $1/day$), $G_o$ is the mean spectral power, $\rho$ is the mean autocorrelation coefficient calculated as $\rho = exp(-\Delta t_{mean}/\tau)$ where $\Delta t_{mean}$ is the arithmetic mean of the time steps in the time series. 
\item Correction factors are calculated for the deviations of the mean artificial periodogram from the analytical value
\begin{equation}
c(f) =<G_{am}(f)>/ G(f)
\end{equation}
These correction factors account for biases resulting from taking periodograms of irregularly spaced data.
\item The unbiased periodogram of the original series is calculated as $G_{ts}(f)/c(f)$, where $G_{ts}(f)$ is the power spectrum of the original time series. The standard deviation is scaled similarly, and plotted in Figure \ref{fig:timeseries}. Assuming a Gaussian distribution, the $95\%$ confidence level is taken to be 2 standard deviations from the mean. The periods of the significant peaks are labeled.
\end{enumerate}
The results are shown in Figs \ref{fig:timeseries} and \ref{fig:timeseries2}.
\subsection{Physical Interpretations of the Oscillations at 283-nm}

\begin{figure*}
\centering
  \includegraphics[width=\linewidth]{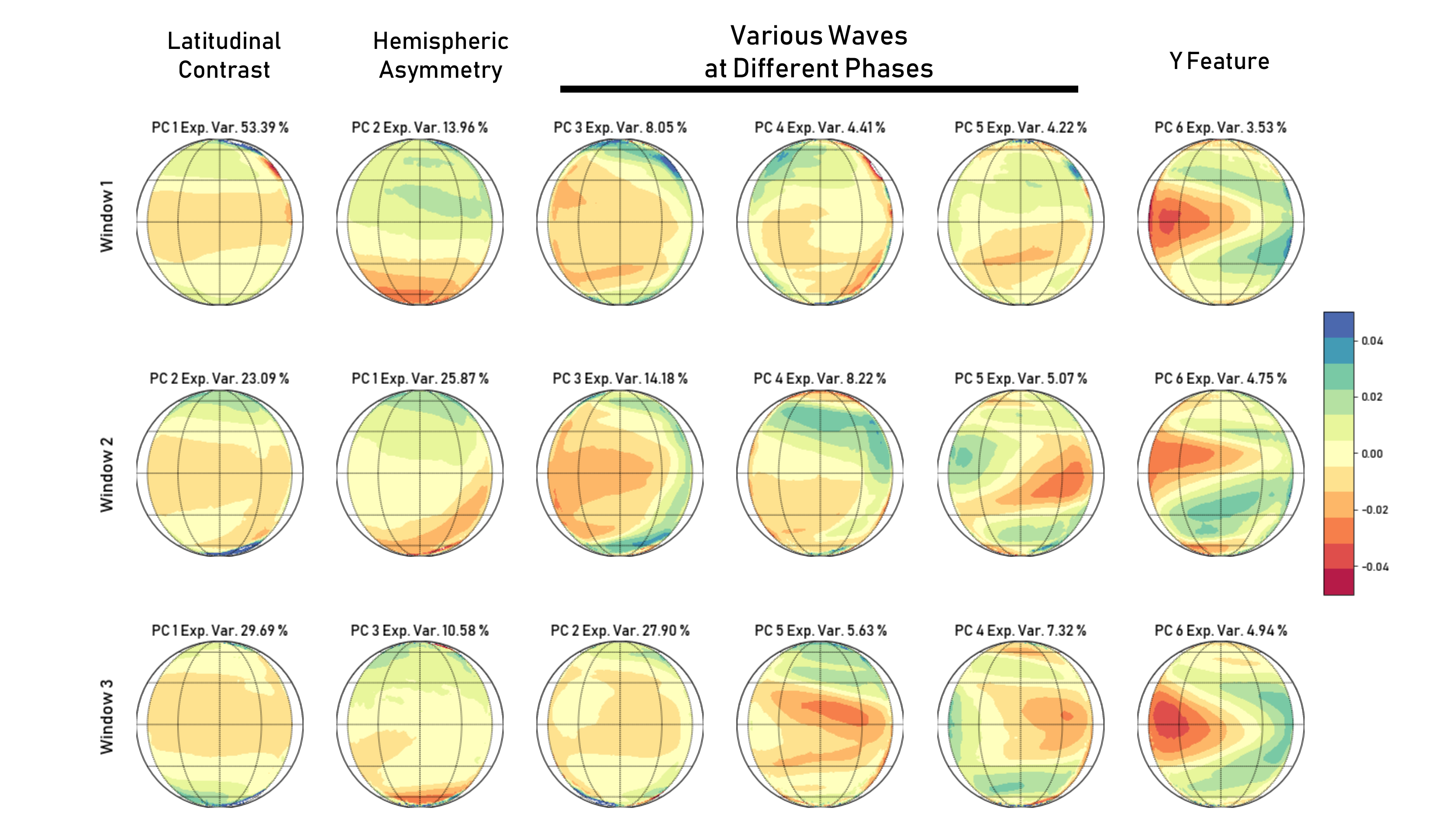}
  \caption{The first six modes of variability from the DINEOF interpolated UVI dataset for the three windows at 283-nm. The first row is lined up in the order of the principal components, while the second and third are rearranged so that similar patterns line up in the same column. The first mode shows a pattern similar to a Hadley circulation, while the second show a hemispheric oscillation. The third, fourth and fifth columns appear to show a combination of short period atmospheric waves while the sixth shows the well known Y-feature.}
  \label{fig:principalcomps}
\end{figure*}

 The first column in Fig \ref{fig:principalcomps} shows a (roughly) longitudinally-uniform latitudinal gradient from the equator to the poles. This likely reflects the Hadley circulation which lifts SO$_2$ from the lower atmosphere to the cloud tops at the equator, from where it is advected to the poles while being lost to photodissociation and conversion to sulphuric acid \citep{marcq2013variations} and polysulphur \citep{mills2007atmospheric}. The corresponding periodicities in Fig \ref{fig:timeseries} show a major periodicity of around 10 days, which is likely due to the spacecraft orbital period \citep{nakamura2016akatsuki}. The synchronization with the orbital motion might indicate a solar-locked component, i.e., a contribution of thermal tides. The approximately four day period can be attributed to the atmospheric rotation around the planet, and the eight day peak is probably a subharmonic of this period. However, GCM studies indicate the existence of barotropic or baroclinic waves with periods ranging from 6-23 days \citep{lebonnois2016wave} in the cloud layer, which is another possibility for the 8 day signature. An $\sim$ 8 day period wave was observed a little deeper than cloud top level in near-infrared observations ($\sim$1.7 micron), and is also thought to result from the interaction of Kelvin and Rossby waves with the mean flow \citep{hosouchi2012wave}.

The second column shows a pattern broadly described as a hemispheric oscillation, showing an oscillation with opposite directions of change between the two polar regions. A polar asymmetric brightening/darkening was observed by a ground-based telescope \citep{dollfus1975venus}, and it was thought to be caused by polar clouds independently evolving in either hemisphere. The pattern identified here is somewhat different, in that the two hemispheres do not evolve independently, rather they evolve together in opposite directions, that is, one brightens while the other darkens. Notably, Fig \ref{fig:timeseries} shows no significant periods associated with this pattern other than the spacecraft orbital period. This can mean either that the phenomenon is aperiodic, or more likely, that the period is longer than the observational window considered here (approximately 40 days). Such a large scale pole-to-pole pattern could indicate the existence of an atmospheric teleconnection. Another interpretation might be that the period comparable to the spacecraft orbit might indicate a contribution of asymmetric components of thermal tides. Alternatively, it may also arise from the non-equatorially symmetric component of the meridional circulation. However, further observations are required for confirmation, as completely symmetric polar features were observed from the mid-infrared ground-based observations \citep{sato2014cloud}. A very recent study of IR1 (900-nm) images from Akatsuki revealed an asymmetric pattern in the middle clouds at low latitudes with a periodicity of 4-5 days \citep{peralta2018morphology}. It is unclear at this time if this IR pattern is directly related the hemispheric mode we find at the cloud tops. Analysis of mid-infrared images  \citep[LIR,][]{fukuhara2011lir} from the same time period as the UV images studied here, using PCA, may confirm this oscillation and offer clues to robust physical interpretations in near future. Notable hemispheric dichotomies have also been observed in CO concentrations below the clouds \citep{arney2014spatially, marcq2006remote, marcq2008latitudinal}. Higher concentrations were found to variably occur either on the Northern and Southern hemispheres during observations separated by several months. SO$_2$ hemispheric dichotomies were also studied, but an unambiguous detection was only made once in 2010. It is unclear at this time how these long-term variations are related to the short-term mode we find.  

 We interpret the third, fourth and fifth columns to be combinations of atmospheric wave patterns. They often show short-period variabilities of approximately 5 days. Waves with periods of a few days have been  observed \citep{del1990planetary,kouyama2015vertical,imai2016ground} and simulated in Venus GCM models \citep{yamamoto2006superrotation}. In particular, the 5 day wave has been interpreted as a Rossby wave and the four day wave as an equatorial Kelvin wave \citep{rossow1990cloud,kouyama2012horizontal,imamura2006meridional}. It must be noted also that several subplots do not show any notable periodicity even similar spatial patterns are associated with waves in other windows. For example, in column three, row one shows no periodicity except for the spacecraft orbit and some high frequency noise around the 1-2 days range. But rows two and three clearly show five and four day periods respectively. It is interesting that similar spatial patterns appear to form from different wave configurations in different periods.
 
 The sixth column clearly shows the famous Y-feature \citep{boyer1961observations,rossow1980cloud}, and is clearly associated with a strong periodicity of about four days. This is also consistent with the interpretation of the major cloud features of the Venus atmosphere interpreted as a trapped Kelvin wave \citep{del1990planetary,kouyama2012horizontal,peralta2015venus}. It is interesting that the order of PCs in each window is slightly variable and needs rearrangement to align similar patterns under each column. This indicates that different processes dominate albedo variability in different periods, which is consistent with the previous finding that the dominant periodicity of atmospheric waves in the Venus atmosphere varies on a timescale of several months \citep{del1990planetary,kouyama2013long,imai2016ground}.
 
 A joint PCA of data in all windows combined together was also attempted. The periodograms of that study were dominated by peaks at about 220 days and its harmonics at 110, 55 etc days, which are functions of the changing orbital orientation of the spacecraft relative to the Venus dayside, the observational window frequency and other factors rather than interesting atmospheric phenomenon. Also, the dominance of these observational periodicities in the data caused peaks from the 4 and 5 day waves to become statistically insignificant. As such, results from that analysis are not discussed in this paper since they are dominated by systematics.
\subsection{Physical Interpretations of the Oscillations at 365-nm}
\begin{figure*}
\centering
  \includegraphics[width=\linewidth]{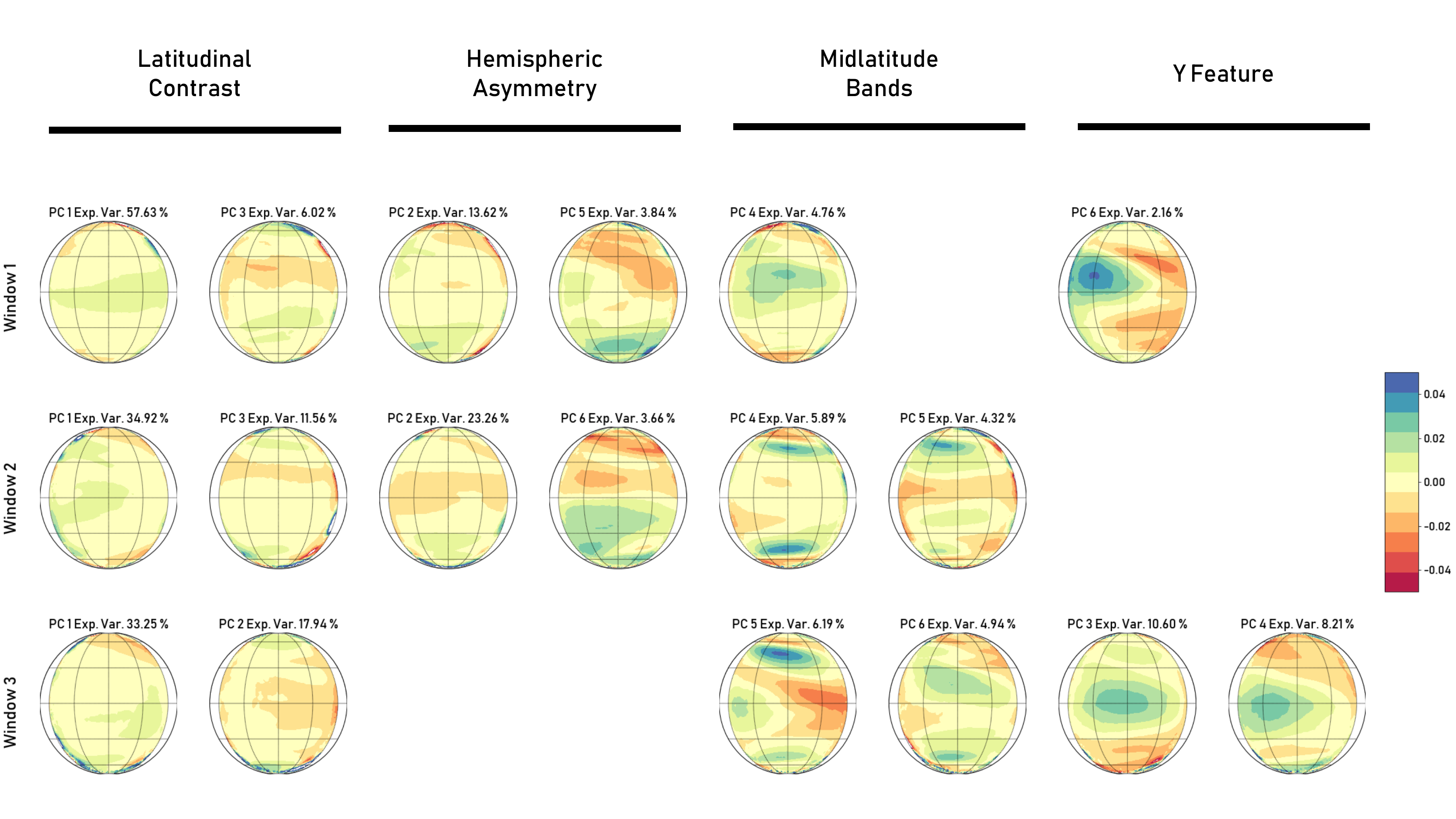}
  \caption{Same as Fig \ref{fig:principalcomps} but for 365-nm. The patterns are much more variable between the windows, but can still be broadly classified into four groups. The first mode shows a pattern similar to a Hadley circulation, while the second shows a hemispheric oscillation. The third and fourth groups show signatures of the midlatitude Rossby waves and the Y-feature respectively. Not all groups are represented in all windows.}
  \label{fig:principalcomps2}
\end{figure*}

\begin{figure*}
\centering
  \includegraphics[width=\linewidth]{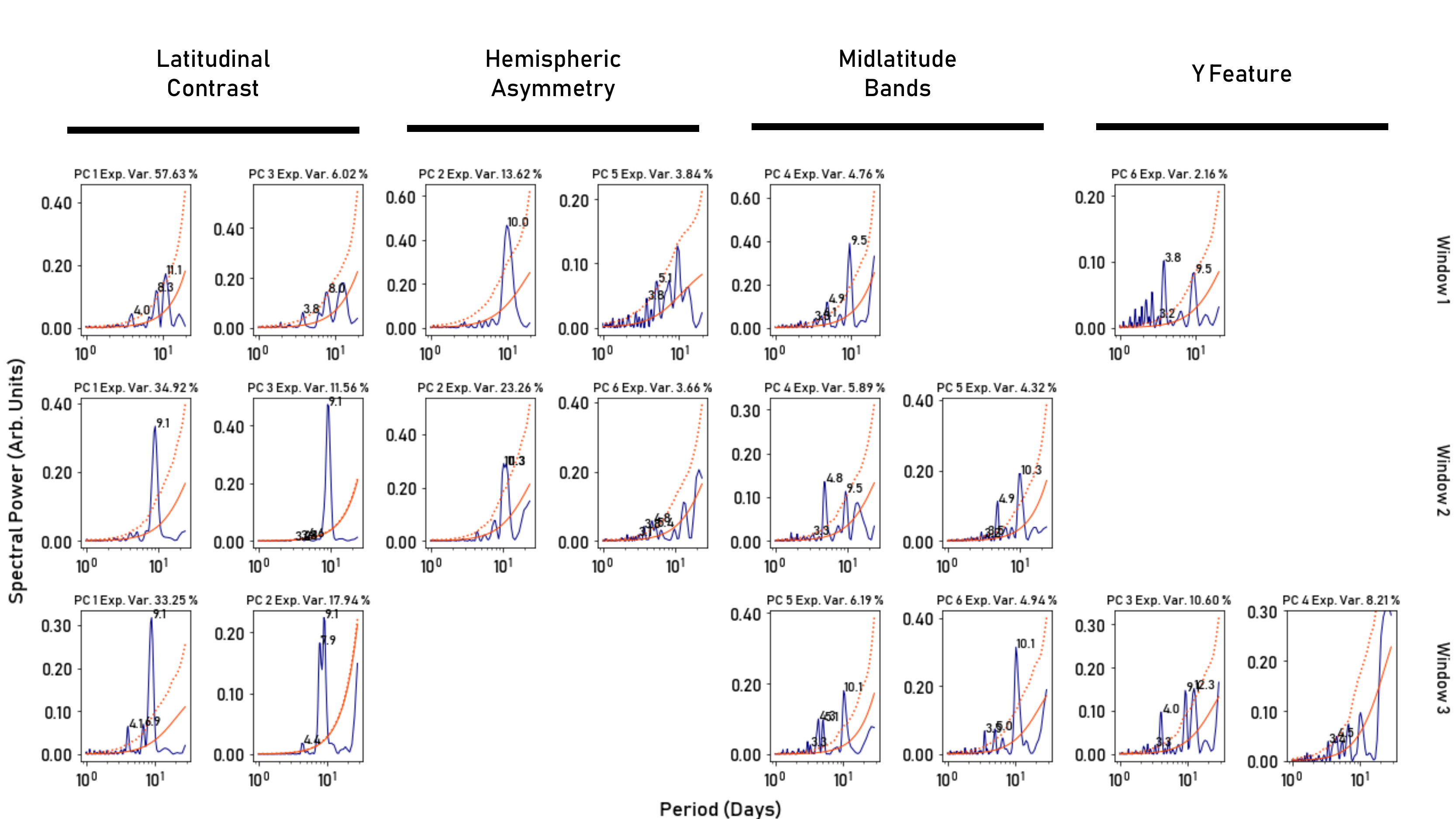}
  \caption{Same as Fig \ref{fig:timeseries} but for 365-nm. }
  \label{fig:timeseries2}
\end{figure*}

The 365-nm PCs are comparatively much more variable across the three windows than the 283-nm PCs as shown in Fig \ref{fig:principalcomps2}. However, they can still be approximately classified into a few broad categories. Two PCs in each window can be placed into the first category, which is similar to the latitudinal gradients seen at 283-nm. The periodicities, as seen in Fig \ref{fig:timeseries2}, are also similar to those seen at 283-nm. This is generally consistent with the understanding that the overturning circulation is responsible for increased UV absorber concentrations leading to low albedos at the equator at 365-nm \citep{titov2008atmospheric, molaverdikhani2012abundance}. However, latitudinal gradients are much weaker than those seen at 283-nm, suggesting that choatic variability in morphology is perhaps more important at this wavelength. Spatial distributions are in general far more complicated at 365-nm, with significant ambiguities in classification. This can be understood on the condition that the 365-nm observations are sensing an altitude level slightly below the cloud top, while 283-nm is at a higher altitude \citep{horinouchi2018mean}. Therefore, the 365-nm features are affected by both the absorber and bright sulfuric acid cloud aerosols, which are formed through photochemical process \citep{mills2007atmospheric,parkinson2015distribution}, whereas the 283-nm would reveal the absorbers above the clouds so atmospheric flow patterns are more apparent. The latter would be also affected by photodissociation of SO$_2$ \citep{mills2007atmospheric} and the upper haze \citep{luginin2016aerosol}, but our results suggest that these influences may be less effective than at 365-nm.

The second category is the hemispheric asymmetry, represented by two PCs each in the first two windows, but appears to be absent in the third. PC 3 in window 3 shows something like a hemispheric mode, however, it is associated with a strong wave peak of around 4 days in Fig \ref{fig:timeseries2}, which is characteristic of the Y-feature. Window 3, PC 4 may also qualify for this category, but is somewhat ambiguous because the pattern is not purely hemispheric. The hemispheric mode typically does not show any strong periodicities apart from the spacecraft orbital period, as seen at 283-nm, though hints of the 4 and 5 day periods are seen when the asymmetry is strong (window 1, PC 5 and window 2, PC 6). The large variability between windows could be attributed to concealment of gradients by transient cloud features, as for the previous category.

The third category shows a very clear transition region around 50$^\circ$ N/S, from dark low-to-middle latitudes to the bright polar hood \citep{titov2012morphology}, sometimes symmetric across both hemispheres. The structure is associated with drastic changes in cloud tracked 365-nm winds, slowing down poleward of about 50$^\circ$ \citep{kouyama2012horizontal}, in cloud top altitudes which also decrease poleward \citep{ignatiev2009altimetry}, and in thermal structure, presenting colder temperatures near the cloud top level \citep{tellmann2009structure}. This means that winds, cloud top structure, and thermal structure are closely linked with the 365-nm patterns, implying overlapped vertical locations each other. The morphology bears a resemblance to model simulations of Rossby wave patterns at the cloud top \citep{kouyama2015vertical}. The periodicities always show a peak at around the 5 day mark confirming the signature of midlatitude Rossby waves. This can be interpreted as features arising from the perturbation of the midlatitude bright bands or the "cold collar" \citep{titov2008atmospheric,titov2012morphology} by Rossby waves. Interestingly, such a clear Rossby wave signature is not apparent in the 283-nm data.

The fourth category is the Y-feature, associated with a period of $\sim 4$ days with some other waves at 4.5, 9 and 12.3 days occasionally contributing. The absence of the Y-feature in window 2 does not mean that this feature did not occur during this time, since it is clearly seen at 283-nm for the same window. It maybe that the feature was sufficiently obscured at 365-nm that it is not represented in the first six PCs considered here.

\section{Conclusions}
We performed a PCA of the Akatsuki UVI 283-nm and 365-nm data taken over about 1.5 years, subdivided into three observational windows each. The first six PCs over each window are considered significant and show similar morphologies at 283-nm, but are more variable at 365-nm. The difference can be understood if the 365-nm observations are sensitive to altitudes below the cloud top which are affected by transient cloud variability, while 283-nm probes the atmosphere above the cloud tops. Additionally, since the unknown UV absorber is the result of unidentified chemical reaction chains, the kinetics of those reactions may also be responsible for some of the observed differences. The signatures of the overturning circulation, Rossby and Kelvin waves are apparent from the spatial patterns and associated periodicities. We also note that similar spatial patterns are sometimes associated with different waves, and the relative importance of different waves change across the different observing periods as seen by the changing order of PCs with similar spatial patterns. 

A hemispheric asymmetry mode is also apparent from this analysis. To be sure of its existence and to understand its dynamics, the same mode will need to be studied in other atmospheric observations of Venus. The analysis technique described here is general, and has potential for use in other gridded atmospheric datasets from Akatsuki, Venus Express and Pioneer Venus, and can be applied to other planetary atmospheric image analysis with a long-term monitoring dataset. 
\section{Acknowledgements}
We thank Takehiko Satoh and the anonymous referee for comments that improved the paper. PK acknowledges generous funding support from the grants-in-aid program of JSPS. Codes used in this study are available on request and Akatsuki data will become publicly available at https://darts.isas.jaxa.jp/pub/pds3/staging/.

\newpage
\bibliographystyle{aasjournal}
\bibliography{venusrefs}

\begin{thebibliography}{}
\expandafter\ifx\csname natexlab\endcsname\relax\def\natexlab#1{#1}\fi
\providecommand{\url}[1]{\href{#1}{#1}}

\bibitem[{Allen \& Smith(1996)}]{allen1996monte}
Allen, M.~R., \& Smith, L.~A. 1996, Journal of climate, 9, 3373

\bibitem[{Alvera-Azc{\'a}rate {et~al.}(2005)Alvera-Azc{\'a}rate, Barth, Rixen,
  \& Beckers}]{alvera2005reconstruction}
Alvera-Azc{\'a}rate, A., Barth, A., Rixen, M., \& Beckers, J.-M. 2005, Ocean
  Modelling, 9, 325

\bibitem[{Arney {et~al.}(2014)Arney, Meadows, Crisp, Schmidt, Bailey, \&
  Robinson}]{arney2014spatially}
Arney, G., Meadows, V., Crisp, D., {et~al.} 2014, Journal of Geophysical
  Research: Planets, 119, 1860

\bibitem[{Beckers \& Rixen(2003)}]{beckers2003eof}
Beckers, J.-M., \& Rixen, M. 2003, Journal of Atmospheric and oceanic
  technology, 20, 1839

\bibitem[{Bickel {et~al.}(2008)Bickel, Levina,
  {et~al.}}]{bickel2008regularized}
Bickel, P.~J., Levina, E., {et~al.} 2008, The Annals of Statistics, 36, 199

\bibitem[{Boyer \& Camichel(1961)}]{boyer1961observations}
Boyer, C., \& Camichel, H. 1961, in Annales d'Astrophysique, Vol.~24, 531

\bibitem[{Burgess \& Webster(1980)}]{burgess1980optimal}
Burgess, T.~M., \& Webster, R. 1980, Journal of soil science, 31, 315

\bibitem[{Cangelosi \& Goriely(2007)}]{cangelosi2007component}
Cangelosi, R., \& Goriely, A. 2007, Biology direct, 2, 2

\bibitem[{Carter(2006)}]{carter2006solutions}
Carter, R.~L. 2006, Research \& Practice in Assessment, 1, 4

\bibitem[{Cattell(1966)}]{cattell1966scree}
Cattell, R.~B. 1966, Multivariate behavioral research, 1, 245

\bibitem[{Del~Genio \& Rossow(1990)}]{del1990planetary}
Del~Genio, A.~D., \& Rossow, W.~B. 1990, Journal of the Atmospheric Sciences,
  47, 293

\bibitem[{Dollfus(1975)}]{dollfus1975venus}
Dollfus, A. 1975, Journal of the Atmospheric Sciences, 32, 1060

\bibitem[{Dray \& Josse(2015)}]{dray2015principal}
Dray, S., \& Josse, J. 2015, Plant Ecology, 216, 657

\bibitem[{Encrenaz {et~al.}(2012)Encrenaz, Greathouse, Roe, Richter, Lacy,
  B{\'e}zard, Fouchet, \& Widemann}]{encrenaz2012hdo}
Encrenaz, T., Greathouse, T., Roe, H., {et~al.} 2012, Astronomy \&
  Astrophysics, 543, A153

\bibitem[{Fukuhara {et~al.}(2011)Fukuhara, Taguchi, Imamura, Nakamura, Ueno,
  Suzuki, Iwagami, Sato, Mitsuyama, Hashimoto, {et~al.}}]{fukuhara2011lir}
Fukuhara, T., Taguchi, M., Imamura, T., {et~al.} 2011, Earth, planets and
  space, 63, 1009

\bibitem[{Fukuhara {et~al.}(2017)Fukuhara, Futaguchi, Hashimoto, Horinouchi,
  Imamura, Iwagaimi, Kouyama, Murakami, Nakamura, Ogohara,
  {et~al.}}]{fukuhara2017large}
Fukuhara, T., Futaguchi, M., Hashimoto, G.~L., {et~al.} 2017, Nature
  Geoscience, 10, 85

\bibitem[{Garcia(2010)}]{garcia2010robust}
Garcia, D. 2010, Computational statistics \& data analysis, 54, 1167

\bibitem[{Gilman {et~al.}(1963)Gilman, Fuglister, \&
  Mitchell~Jr}]{gilman1963power}
Gilman, D.~L., Fuglister, F.~J., \& Mitchell~Jr, J.~M. 1963, Journal of the
  Atmospheric Sciences, 20, 182

\bibitem[{Horinouchi {et~al.}(2018)Horinouchi, Kouyama, Lee, Murakami, Ogohara,
  Takagi, Imamura, Nakajima, Peralta, Yamazaki, {et~al.}}]{horinouchi2018mean}
Horinouchi, T., Kouyama, T., Lee, Y.~J., {et~al.} 2018, Earth, Planets and
  Space, 70, 10

\bibitem[{Hosouchi {et~al.}(2012)Hosouchi, Kouyama, Iwagami, Ohtsuki, \&
  Takagi}]{hosouchi2012wave}
Hosouchi, M., Kouyama, T., Iwagami, N., Ohtsuki, S., \& Takagi, M. 2012,
  Icarus, 220, 552

\bibitem[{Ignatiev {et~al.}(2009)Ignatiev, Titov, Piccioni, Drossart,
  Markiewicz, Cottini, Roatsch, Almeida, \& Manoel}]{ignatiev2009altimetry}
Ignatiev, N., Titov, D., Piccioni, G., {et~al.} 2009, Journal of Geophysical
  Research: Planets, 114

\bibitem[{Ilin \& Raiko(2010)}]{ilin2010practical}
Ilin, A., \& Raiko, T. 2010, Journal of Machine Learning Research, 11, 1957

\bibitem[{Imai {et~al.}(2016)Imai, Takahashi, Watanabe, Kouyama, Watanabe,
  Gouda, \& Gouda}]{imai2016ground}
Imai, M., Takahashi, Y., Watanabe, M., {et~al.} 2016, Icarus, 278, 204

\bibitem[{Imamura(2006)}]{imamura2006meridional}
Imamura, T. 2006, Journal of the atmospheric sciences, 63, 1623

\bibitem[{Jackson(1993)}]{jackson1993stopping}
Jackson, D.~A. 1993, Ecology, 74, 2204

\bibitem[{Khatuntsev {et~al.}(2013)Khatuntsev, Patsaeva, Titov, Ignatiev,
  Turin, Limaye, Markiewicz, Almeida, Roatsch, \& Moissl}]{khatuntsev2013cloud}
Khatuntsev, I., Patsaeva, M., Titov, D., {et~al.} 2013, Icarus, 226, 140

\bibitem[{Kondrashov \& Ghil(2006)}]{kondrashov2006spatio}
Kondrashov, D., \& Ghil, M. 2006, Nonlinear Processes in Geophysics, 13, 151

\bibitem[{Kouyama {et~al.}(2012)Kouyama, Imamura, Nakamura, Satoh, \&
  Futaana}]{kouyama2012horizontal}
Kouyama, T., Imamura, T., Nakamura, M., Satoh, T., \& Futaana, Y. 2012,
  Planetary and Space Science, 60, 207

\bibitem[{Kouyama {et~al.}(2013)Kouyama, Imamura, Nakamura, Satoh, \&
  Futaana}]{kouyama2013long}
---. 2013, Journal of Geophysical Research: Planets, 118, 37

\bibitem[{Kouyama {et~al.}(2015)Kouyama, Imamura, Nakamura, Satoh, \&
  Futaana}]{kouyama2015vertical}
---. 2015, Icarus, 248, 560

\bibitem[{Lebonnois {et~al.}(2016)Lebonnois, Sugimoto, \&
  Gilli}]{lebonnois2016wave}
Lebonnois, S., Sugimoto, N., \& Gilli, G. 2016, Icarus, 278, 38

\bibitem[{Lee {et~al.}(2015)Lee, Imamura, Schr{\"o}der, \& Marcq}]{lee2015long}
Lee, Y., Imamura, T., Schr{\"o}der, S., \& Marcq, E. 2015, Icarus, 253, 1

\bibitem[{Lee {et~al.}(2017)Lee, Yamazaki, Imamura, Yamada, Watanabe, Sato,
  Ogohara, Hashimoto, \& Murakami}]{lee2017scattering}
Lee, Y., Yamazaki, A., Imamura, T., {et~al.} 2017, The Astronomical Journal,
  154, 44

\bibitem[{Luginin {et~al.}(2016)Luginin, Fedorova, Belyaev, Montmessin,
  Wilquet, Korablev, Bertaux, \& Vandaele}]{luginin2016aerosol}
Luginin, M., Fedorova, A., Belyaev, D., {et~al.} 2016, Icarus, 277, 154

\bibitem[{Marcq {et~al.}(2013)Marcq, Bertaux, Montmessin, \&
  Belyaev}]{marcq2013variations}
Marcq, E., Bertaux, J.-L., Montmessin, F., \& Belyaev, D. 2013, Nature
  geoscience, 6, 25

\bibitem[{Marcq {et~al.}(2008)Marcq, B{\'e}zard, Drossart, Piccioni, Reess, \&
  Henry}]{marcq2008latitudinal}
Marcq, E., B{\'e}zard, B., Drossart, P., {et~al.} 2008, Journal of Geophysical
  Research: Planets, 113

\bibitem[{Marcq {et~al.}(2006)Marcq, Encrenaz, B{\'e}zard, \&
  Birlan}]{marcq2006remote}
Marcq, E., Encrenaz, T., B{\'e}zard, B., \& Birlan, M. 2006, Planetary and
  Space Science, 54, 1360

\bibitem[{Markiewicz {et~al.}(2007)Markiewicz, Titov, Limaye, Keller, Ignatiev,
  Jaumann, Thomas, Michalik, Moissl, \& Russo}]{markiewicz2007morphology}
Markiewicz, W., Titov, D., Limaye, S., {et~al.} 2007, Nature, 450, 633

\bibitem[{Meinke {et~al.}(2005)Meinke, DeVoil, Hammer, Power, Allan, Stone,
  Folland, \& Potgieter}]{meinke2005rainfall}
Meinke, H., DeVoil, P., Hammer, G.~L., {et~al.} 2005, Journal of Climate, 18,
  89

\bibitem[{Mills {et~al.}(2007)Mills, Esposito, \& Yung}]{mills2007atmospheric}
Mills, F.~P., Esposito, L.~W., \& Yung, Y.~L. 2007, Geophysical Monograph
  Series

\bibitem[{Molaverdikhani {et~al.}(2012)Molaverdikhani, McGouldrick, \&
  Esposito}]{molaverdikhani2012abundance}
Molaverdikhani, K., McGouldrick, K., \& Esposito, L.~W. 2012, Icarus, 217, 648

\bibitem[{Mudelsee(2002)}]{mudelsee2002tauest}
Mudelsee, M. 2002, Computers \& Geosciences, 28, 69

\bibitem[{Nakagawa \& Freckleton(2008)}]{nakagawa2008missing}
Nakagawa, S., \& Freckleton, R.~P. 2008, Trends in Ecology \& Evolution, 23,
  592

\bibitem[{Nakamura {et~al.}(2016)Nakamura, Imamura, Ishii, Abe, Kawakatsu,
  Hirose, Satoh, Suzuki, Ueno, Yamazaki, {et~al.}}]{nakamura2016akatsuki}
Nakamura, M., Imamura, T., Ishii, N., {et~al.} 2016, Earth, Planets and Space,
  68, 75

\bibitem[{Ogohara {et~al.}(2017)Ogohara, Takagi, Murakami, Horinouchi, Yamada,
  Kouyama, Hashimoto, Imamura, Yamamoto, Kashimura,
  {et~al.}}]{ogohara2017overview}
Ogohara, K., Takagi, M., Murakami, S.-y., {et~al.} 2017, Earth, Planets and
  Space, 69, 167

\bibitem[{Parkinson {et~al.}(2015)Parkinson, Gao, Schulte, Bougher, Yung,
  Bardeen, Wilquet, Vandaele, Mahieux, Tellmann,
  {et~al.}}]{parkinson2015distribution}
Parkinson, C.~D., Gao, P., Schulte, R., {et~al.} 2015, Planetary and Space
  Science, 113, 205

\bibitem[{Peralta {et~al.}(2015)Peralta, S{\'a}nchez-Lavega,
  L{\'o}pez-Valverde, Luz, \& Machado}]{peralta2015venus}
Peralta, J., S{\'a}nchez-Lavega, A., L{\'o}pez-Valverde, M., Luz, D., \&
  Machado, P. 2015, Geophysical Research Letters, 42, 705

\bibitem[{Peralta {et~al.}(2018)Peralta, Iwagami, S{\'a}nchez-Lavega, Lee,
  Hueso, Narita, Imamura, Miles, Wesley, \& Kardasis}]{peralta2018morphology}
Peralta, J., Iwagami, N., S{\'a}nchez-Lavega, A., {et~al.} 2018, Geophysical
  Research Letters, 46, 1

\bibitem[{Peres-Neto {et~al.}(2005)Peres-Neto, Jackson, \&
  Somers}]{peres2005many}
Peres-Neto, P.~R., Jackson, D.~A., \& Somers, K.~M. 2005, Computational
  Statistics \& Data Analysis, 49, 974

\bibitem[{Pourahmadi(2011)}]{pourahmadi2011covariance}
Pourahmadi, M. 2011, Statistical Science, 369

\bibitem[{Ross(1928)}]{ross1928photographs}
Ross, F.~E. 1928, The Astrophysical Journal, 68, 57

\bibitem[{Rossow {et~al.}(1990)Rossow, Del~Genio, \& Eichler}]{rossow1990cloud}
Rossow, W.~B., Del~Genio, A.~D., \& Eichler, T. 1990, Journal of the
  Atmospheric Sciences, 47, 2053

\bibitem[{Rossow {et~al.}(1980)Rossow, Del~Genio, Limaye, Travis, \&
  Stone}]{rossow1980cloud}
Rossow, W.~B., Del~Genio, A.~D., Limaye, S.~S., Travis, L.~D., \& Stone, P.~H.
  1980, Journal of Geophysical Research: Space Physics, 85, 8107

\bibitem[{Sato {et~al.}(2014)Sato, Sagawa, Kouyama, Mitsuyama, Satoh, Ohtsuki,
  Ueno, Kasaba, Nakamura, \& Imamura}]{sato2014cloud}
Sato, T.~M., Sagawa, H., Kouyama, T., {et~al.} 2014, Icarus, 243, 386

\bibitem[{Schneider(2001)}]{schneider2001analysis}
Schneider, T. 2001, Journal of climate, 14, 853

\bibitem[{Schulz \& Mudelsee(2002)}]{schulz2002redfit}
Schulz, M., \& Mudelsee, M. 2002, Computers \& Geosciences, 28, 421

\bibitem[{Tellmann {et~al.}(2009)Tellmann, P{\"a}tzold, H{\"a}usler, Bird, \&
  Tyler}]{tellmann2009structure}
Tellmann, S., P{\"a}tzold, M., H{\"a}usler, B., Bird, M.~K., \& Tyler, G.~L.
  2009, Journal of Geophysical Research: Planets, 114

\bibitem[{Titov {et~al.}(2008)Titov, Taylor, Svedhem, Ignatiev, Markiewicz,
  Piccioni, \& Drossart}]{titov2008atmospheric}
Titov, D.~V., Taylor, F.~W., Svedhem, H., {et~al.} 2008, Nature, 456, 620

\bibitem[{Titov {et~al.}(2012)Titov, Markiewicz, Ignatiev, Song, Limaye,
  Sanchez-Lavega, Hesemann, Almeida, Roatsch, Matz,
  {et~al.}}]{titov2012morphology}
Titov, D.~V., Markiewicz, W.~J., Ignatiev, N.~I., {et~al.} 2012, Icarus, 217,
  682

\bibitem[{Wang {et~al.}(2012)Wang, Garcia, Liu, De~Jeu, \&
  Dolman}]{wang2012three}
Wang, G., Garcia, D., Liu, Y., De~Jeu, R., \& Dolman, A.~J. 2012, Environmental
  Modelling \& Software, 30, 139

\bibitem[{Warton(2008)}]{warton2008penalized}
Warton, D.~I. 2008, Journal of the American Statistical Association, 103, 340

\bibitem[{Wilks(2006)}]{wilks2006statistical}
Wilks, D.~S. 2006, Statistical Methods in the Atmospheric Sciences
  (International Geophysics Series; V. 91) (Academic Press)

\bibitem[{Yamamoto \& Takahashi(2006)}]{yamamoto2006superrotation}
Yamamoto, M., \& Takahashi, M. 2006, Journal of the atmospheric sciences, 63,
  3296

\bibitem[{Yamazaki {et~al.}(2018)Yamazaki, Yamada, Lee, Watanabe, Horinouchi,
  Murakami, Kouyama, Ogohara, Imamura, Sato,
  {et~al.}}]{yamazaki2018ultraviolet}
Yamazaki, A., Yamada, M., Lee, Y.~J., {et~al.} 2018, Earth, Planets and Space,
  70, 23

\end{thebibliography}
\end{document}